\documentclass[
  aps,prl,reprint,nofootinbib,longbibliography,
  amsmath,amssymb
]{revtex4-2}

\usepackage[colorlinks=true]{hyperref}
\usepackage{graphicx}% Include figure files
\usepackage{dcolumn}% Align table columns on decimal point
\usepackage{bm}% bold math
\usepackage{mathrsfs}

\usepackage{tikz}
\usetikzlibrary{arrows.meta,calc}
\usetikzlibrary{decorations.pathmorphing, positioning}

\usepackage[utf8]{inputenc}
\usepackage[greek,english]{babel}
\usepackage{lmodern} % or cm-super for more Unicode coverage

\usepackage[T1]{fontenc}
\usepackage[utf8]{inputenc}
\usepackage{bm}
\usepackage{amsmath,amssymb}
\usepackage{dsfont}
\usepackage{comment}
\usepackage{mathtools}

\newcommand{\ket}[1]{\lvert #1\rangle}
\newcommand{\bra}[1]{\langle #1\rvert}

\newcommand{\ketbra}[2]{\lvert #1\rangle\!\langle #2\rvert}
\newcommand{\tr}{\mathrm{tr}}
\newcommand{\one}{\mathds{1}}

\usepackage{makecell}

\newif\ifcombined
\newcommand{\SMref}[2]{\ifcombined\ref{#1}\else#2\fi}

\newcommand{\SMreferences}{%
  \nocite{bedard2021abc,banaszek2001fidelity,cerf2000pauli,
    fiurasek2005asymmetric,kretschmann2008continuity}%
}

\newcommand{\BeginSupplement}{%
  \clearpage
  \onecolumngrid

  \setcounter{page}{1}
  \renewcommand{\thepage}{S\arabic{page}}

  \setcounter{figure}{0}
  \renewcommand{\thefigure}{S\arabic{figure}}
  \renewcommand{\theHfigure}{supp.\arabic{figure}}

  \setcounter{table}{0}
  \renewcommand{\thetable}{S\arabic{table}}
  \renewcommand{\theHtable}{supp.\arabic{table}}

  \appendix
  \setcounter{secnumdepth}{2}

  \begin{center}
    {\large\bfseries Supplemental Material}\par\medskip
    {\bfseries Classical Gravity Cannot Locally Mediate Quantum Teleportation}\par\medskip
    Charles Alexandre B\'edard, Simone Rijavec, and Maria Violaris
  \end{center}
  \medskip
}
\combinedtrue

\begin{document}

\title{Classical Gravity Cannot Locally Mediate Quantum Teleportation}% Force line breaks with \\

\author{Charles Alexandre Bédard$^{*}$}
\affiliation{École de technologie supérieure,
Montréal, Québec, Canada}%
\author{Simone Rijavec$^{*}$}
%\thanks{All authors contributed equally to this work.}
\affiliation{School of Physics and Astronomy, Tel-Aviv University, Tel-Aviv 69978, Israel}
\author{Maria Violaris$^{*}$}
%\thanks{All authors contributed equally to this work.}
\affiliation{Department of Physics, University of Oxford, Oxford, United Kingdom}

\date{November 26, 2026}% It is always \today, today,

\begin{abstract}
Quantum teleportation seems to live up to its name: shared entanglement and two classical bits suffice to transfer an arbitrary qubit state to a distant location. What, if anything, physically carries the continuous parameters specifying that state? We formulate the principle of locally mediated information and prove that, together with no-cloning, it forbids perfect teleportation through a fundamentally classical medium. 
This turns teleportation into a probe of the medium’s fundamental nature. 
%This makes the fundamental nature of the communication channel an experimental question.
%We propose putting gravity to this test by using it to transmit the two bits. 
We propose using gravity to transmit the two bits.
Under local mediation, perfect gravitational teleportation would witness gravity’s nonclassicality without requiring coherent control of the source masses.
\end{abstract}

\maketitle

 \begingroup
 \renewcommand{\thefootnote}{*}
\footnotetext{\hspace{-4pt}All authors contributed equally to this work.}
 \endgroup

%V9.2 corresponds to V9 on 17/11/27 at 15:58. It is the one shared to Logan. 

In quantum teleportation~\cite{bennett1993teleporting}, Alice holds one share of~$\ket{\Phi^+}$ and another qubit in the state
\begin{equation}
\ket{\psi} = \alpha\ket{0} + \beta\ket{1} \,,
\end{equation} where $\psi \equiv \alpha/\beta
\in \mathbb C \cup \{\infty\} \simeq S^2$ labels a point on the Bloch sphere.
Alice measures her two qubits in the Bell basis and sends the two-bit classical outcome~$(i,j)$ to Bob, who then applies the appropriate Pauli correction to his share of~$\ket{\Phi^+}$.
This brings it into the state~$\ket{\psi}$.
Penrose~\cite{penrose1998quantum} raised the problem: ``How is it that the continuous `information' of the spin direction of the state that she wishes to transmit [...] can be transmitted to Bob when she actually sends him only two bits of discrete information?''
%Penrose~\cite{penrose1998quantum} raised the problem of how two classical bits, together with shared entanglement, suffice to transfer an arbitrary qubit state.

\begin{figure}[t]
  \centering
  \includegraphics[width=\columnwidth, trim=12cm 5.5cm 12cm 9.7cm, clip]{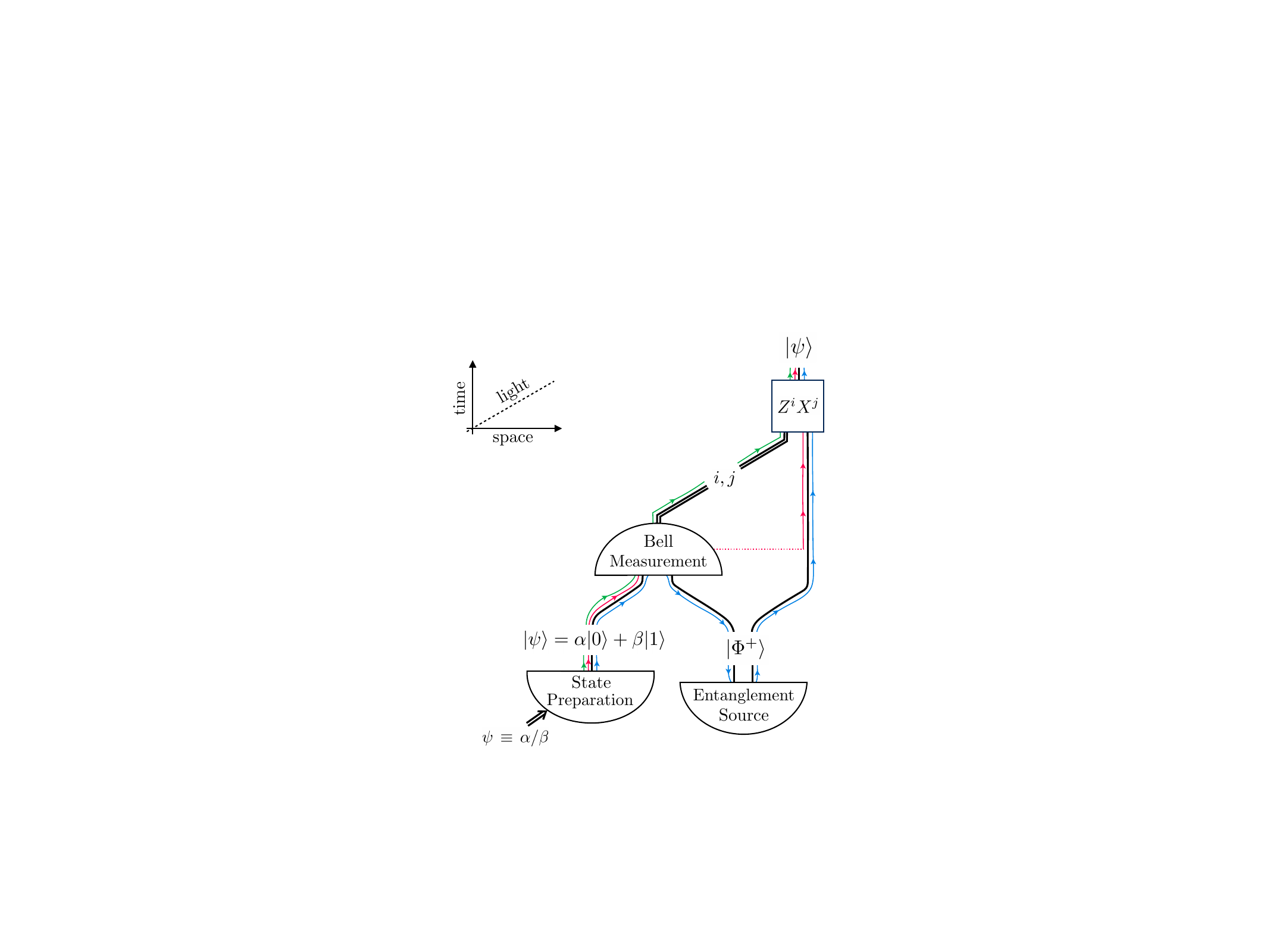}
  \caption{Schematic spacetime representation of quantum teleportation showing proposed routes for the protocol’s \mbox{$\psi$-dependence}: retrocausal (blue), nonlocal (red), and locally mediated (green).
}
  \label{fig:tele}
\end{figure}

Taking the two-bit limitation as decisive, Penrose turned to entanglement as the missing link: the \(\psi\)-dependence would travel backward in time from Alice's measurement to the entanglement source, then forward to Bob. %~\cite{jozsa2004illustrating}.
Other accounts take the ``tele’’ in teleportation quite literally, attributing the state transfer to nonlocality—through textbook or dynamical collapse~\cite{griffiths2018quantum,Zeilinger2007Teleportation,grw1986,
bassi2013}, Bohmian guidance~\cite{maroney1999quantum}, or ``world nonlocality’’~\cite{vaidman2022transfer}.
Some fully unitary analyses in the Schr\"odinger picture instead assign the \mbox{$\psi$-dependence} to global properties of the quantum state~\cite{braunstein1996quantum, timpson2006grammar}, while epistemic approaches deflate the problem by denying that the quantum state is an observer-independent physical entity~\cite{fuchs2019qbism}. 
%whose transmission requires a spacetime carrier

Deutsch and Hayden formulated quantum computation in the Heisenberg picture of unitary quantum theory and analyzed quantum teleportation in those terms~\cite{deutsch2000information, bedard2023teleportation}.
There, the classical bits transport the \mbox{$\psi$-dependence}, locally and causally.
This is possible because the physical systems constituting the communication channel are taken to be quantum systems~\cite{everett1973universal} and thereby admit an operator-valued structure.
Penrose's apparent capacity obstruction is thus overcome: the \(\psi\)-dependence propagates locally through the quantum systems carrying the two classical bits.
Their apparent classicality emerges from decohering interactions with the environment, which stabilize macroscopic records without eliminating the underlying \(\psi\)-dependence. 
That dependence propagates through chains of local interactions, just as ordinary classical messages do.
This is fully consistent with all familiar communication media---air molecules, electronic circuits, electromagnetic signals---which are treated as classical in practice while operating quantum-mechanically at the bottom~(Fig.~\ref{fig:tele}).
%
%Although no measurement on the channel alone reveals~$\psi$, its underlying quantum systems nevertheless carry the $\psi$-dependence, which becomes manifest in Bob's local measurement statistics after his Pauli corrections.
%or : Although~$\psi$ is not directly accessible from the communication channel alone, as per that account it nevertheless resides in the channel’s degrees of freedom and is rendered accessible again upon Bob's Pauli corrections.

These accounts disagree profoundly about how teleportation works, yet standard experiments have not distinguished them.
The Deutsch–Hayden mechanism suggests where an empirical difference might arise.
What if the systems constituting the classical channel were \emph{fundamentally} classical?

%In this work, we show that this question makes the locality assumptions underlying the Deutsch–Hayden account experimentally testable. This becomes particularly relevant in the case of gravity, whose fundamental nature remains unsettled. We thus propose a new experiment, gravitational teleportation, whose successful implementation would witness the nonclassicality of gravity under our locality assumptions, or challenge these assumptions if gravity were independently established to be fundamentally classical.

To investigate this possibility, we abstract from the Deutsch–Hayden account its local, mechanistic picture of information propagation, without assuming quantum dynamics for the mediating systems.
Suppose a localized physical system is prepared in a way that depends on a parameter~$\theta$. In a local physical theory, the resulting~$\theta$-dependence does not disappear from one location only to reappear elsewhere. Rather, it propagates through local interactions to neighbouring systems, and from them to others in turn.
We also require that, given enough independent runs, an observer can reliably estimate~\(\theta\) to any desired accuracy from the systems reached by these interactions.
%We require more than the local propagation of~\(\theta\)-dependence:~\(\theta\) must remain recoverable from the collection of systems reached through these interactions.

We call this combination of local propagation and tomographic recoverability the principle of locally mediated information (LMI).
Our main result is that, under LMI and no-cloning, a fundamentally classical communication channel—one whose complete local description is arbitrarily copyable—cannot support perfect teleportation.

Formally, LMI imposes two conditions on a physical theory.
The first is the existence of an \emph{Einstein-local description} of composite systems. 
At each time~$t$, every system~$X$ has a complete local description~$s_X^{(t)}$, and the complete description of a composite system~$X_1\cdots X_n$ is given by the collection
$\bigl(s_{X_1}^{(t)},\ldots,s_{X_n}^{(t)}\bigr)$.
In particular, even entangled systems are fully described in terms of the local descriptions of their subsystems.
Moreover, an interaction may change only the local descriptions of the interacting subsystems, in accordance with no action at a distance.

The second condition is \emph{tomographic recoverability}. 
Suppose information about a parameter~$\theta$ is initially instantiated in a physical system. For every time~\mbox{$t\geq0$,} let~$S_t$ comprise all systems reached from the initial system through chains of local interactions, whether or not their individual descriptions depend explicitly on~$\theta$. Tomographic recoverability requires~$\theta$ to remain recoverable from~$S_t$.
More precisely, given~$N$ independent repetitions of operationally equivalent preparation-and-evolution procedures, producing collections~$S_{t,k}(\theta)$ for~$k=1,\ldots,N$, there must exist a recovery task allowed by the theory,
\[
\mathsf{Rec}_{S_t}^N:
\bigl(S_{t,1}(\theta),\ldots,S_{t,N}(\theta)\bigr)
\longrightarrow \widehat{\theta}_N,
\]
where~$\widehat{\theta}_N$ is an estimate of~$\theta$. 
For an appropriate distance~$d$ on the parameter space, tomographic recoverability requires that, for every~$\varepsilon>0$, the estimate can be made to satisfy~$d \bigl(\widehat{\theta}_N,\theta\bigr)<\varepsilon$ with arbitrarily high reliability by taking~$N$ sufficiently large.
Operational equivalence here means identical statistics for every allowed measurement procedure on \(S_t\), not necessarily identical complete local descriptions.
In particular, we assume that qubits in the same pure state $\ket{\psi}$ are operationally interchangeable regardless of their preparation histories. 
%We assume that, for each fixed \(\psi\), a qubit in the pure state \(|\psi\rangle\) is operationally interchangeable with an independently prepared qubit in that state, even in continuations that include joint measurements on systems retained from earlier stages.
%, even in joint experiments with systems they have previously interacted with.
%We take qubits in the same pure state \(\ket{\psi}\) to be operationally interchangeable, regardless of their preparation history, even in joint experiments involving systems with which they have previously interacted. In ordinary quantum theory, this follows from the fact that a subsystem in a pure state cannot be correlated with other systems; here, we assume the same property for hybrid quantum–classical theories.
%
The recovery task need only be physically allowed by the theory; no particular implementation is assumed.

The Deutsch–Hayden formulation realizes LMI within unitary quantum theory through Heisenberg-picture descriptors—the time-evolved local generators of the algebra of observables.
Whether these descriptors represent physical properties has been debated~\cite{timpson2005nonlocality,wallace2007locality,
deutsch_vindication_2012, bedard2026realism}; here, LMI is treated as a physical hypothesis.
 
In an LMI account of teleportation, the relevant parameter is~\mbox{$\theta=\psi\in S^2$}, the Bloch-vector parameter labelling the input qubit state~$\ket{\psi}$.
The $\psi$-dependence is initially localized in Alice's input qubit~$A_1$ and reaches Bob's qubit~$B$ when teleportation is completed.
Between these endpoints, LMI requires the $\psi$-dependence to propagate through chains of local interactions.
Accordingly, \(S_t\) grows to include measurement devices, physical records, systems along the communication channel, and environmental degrees of freedom reached through these interactions.

Consider a collection \(C\) of physical systems constituting a section of the communication channel. We call this collection the \emph{mediator}; its constituent systems may be spatially distributed. Let \(D_C\) denote the systems needed to complete teleportation downstream of \(C\), apart from Bob’s qubit \(B\), including the remainder of the channel and the apparatus implementing the conditional Pauli correction. There is a protocol stage \(t_c\) at which the \(\psi\)-dependence has reached \(C\), but neither \(D_C\) nor \(B\). At this stage, \(C\subseteq S_{t_c}\), and LMI requires \(\psi\) to remain tomographically recoverable from \(S_{t_c}\). We choose \(C\) so that teleportation can then be completed using \(CD_CB\) alone, without further input from upstream systems.
%
%The feed-forward structure of teleportation allows the protocol to be completed without further input from systems upstream of a suitable channel section.
%
%We take~\(C\) to be such a section, enlarging it if necessary to include every physical carrier linking the upstream and downstream portions of the protocol whose local description is required for its continuation.
%
%We call such a section \emph{downstream-sufficient}.
Formally,
\begin{equation}\label{eq:down}
\left(s_C^{(t_c)}(\psi),\,s_{D_C}^{(t_c)},\,s_B^{(t_c)}\right)
\longmapsto s_B^{(t_f)}(\psi),
\end{equation}
where only Bob's final local description is displayed, and~$s_B^{(t_f)}(\psi)$ yields the output state~$\ket{\psi}$.

%The section \(C\subseteq S_{t_c}\) may comprise several spatially distributed systems.
%
%Let~$D_C$ denote the systems required, in addition to Bob's qubit~$B$, to complete the protocol downstream of~$C$.
%
%These may include the remainder of the channel and the apparatus that reads the signal and implements the conditional Pauli correction.
%
%At~$t_c$, the $\psi$-dependence has not yet reached either~$D_C$ or~$B$, while LMI requires~$\psi$ to remain tomographically recoverable from~$S_{t_c}$ alone.

%The feed-forward structure of teleportation allows the protocol to be completed without further input from the systems upstream of a suitable channel section.
%
%We take~$C$ to include every physical carrier crossing the cut whose local description is required for this continuation, enlarging~$C$ if necessary.
%
%We call such a section \emph{downstream-sufficient}: after stage~$t_c$, the local descriptions of~$C$,~$D_C$, and~$B$ suffice to complete teleportation, independently of the remainder of~$S_{t_c}$.
%
%More precisely,
%\begin{equation}\label{eq:down}
%\left(s_C^{(t_c)}(\psi),\,s_{D_C}^{(t_c)},\,s_B^{(t_c)}\right)
%\longmapsto s_B^{(t_f)}(\psi),
%\end{equation}
%where only Bob's final local description is displayed, and~$s_B^{(t_f)}(\psi)$ yields the output state~$\ket{\psi}$.

We now ask whether the mediator~$C$ can be fundamentally classical; no classicality assumption is made about any other system.
We take universal copyability of the entire local description of \(C\) to be a necessary condition for its fundamental classicality.
Let~$C'$ be an isomorphic system initialized in a fixed blank state~$s_{C'}^0$.
For every allowed local description~$s_C$, the task
\[
\mathsf{Copy}_C:\quad
     \left( s_C^{\vphantom{0}},\,s_{C'}^{0} \right)
    \longmapsto
    \left(s_C^{\vphantom{0}},\,s_{C'}^{\vphantom{0}}\right)\,, \qquad s_{C'}^{\vphantom{0}}=s_C^{\vphantom{0}} \,,
\]
must therefore be possible to arbitrary operational accuracy, meaning that, for any finite allowed continuation, the resulting predictions can be made arbitrarily close to those obtained if the displayed copying transformation were exact.
This requirement concerns the complete local description of~$C$, including any $\psi$-dependence it carries, not merely the accessible two-bit signal.
Such copyability, even after interactions with quantum systems, is a substantive assumption; our theorem constrains theories admitting it without establishing that consistent hybrid dynamics realizing it exist.

We also assume the \emph{no-cloning principle} for the qubits used in teleportation: from a single unknown input in an arbitrary state \(\ket{\psi}\), no physical procedure can produce arbitrarily many qubits whose states approach \(\ket{\psi}\) with arbitrarily high fidelity~\cite{wootters1982,dieks1982,keyl1999}.

\textbf{Theorem.} 
Under LMI and the no-cloning principle stated above, perfect teleportation of an arbitrary input state through the mediator \(C\) implies that \(C\) cannot be fundamentally classical.

\emph{Proof.}
Suppose, for contradiction, that perfect teleportation succeeds under LMI through the mediator~\(C\), assumed to be fundamentally classical. 
At protocol stage~$t_c$, the relevant local descriptions can be grouped as
\[
    \left(s_{S_{t_c}}^{(t_c)}(\psi),\,s_{D_C}^{(t_c)},\,s_B^{(t_c)}\right)\,, \qquad C \subseteq S_{t_c} \,.
\]
Since~$C$ is fundamentally classical, $\mathsf{Copy}_C$ can be applied at~$t_c$ to~$C$ and an initially blank system~$C'$.
Suppressing copying errors that can be made arbitrarily small, the post-copy descriptions are
\[
    \left(s_{S_{t_c}}^{(t_c)}(\psi),\, s_{C'}^{(t_c)}(\psi) ,\,s_{D_C}^{(t_c)},\,s_B^{(t_c)}\right)\,,
\]
where~$s_{C'}^{(t_c)}(\psi)=s_C^{(t_c)}(\psi)$.
The original mediator~$C$ remains unchanged as part of~$S_{t_c}$, while~$C'$ is used in the downstream continuation of the protocol. 
By the Einstein-locality condition of LMI, the matching local descriptions of~$C$ and~$C'$ make~$C'D_CB$ and~$CD_CB$ equivalent as inputs to the downstream dynamics.
Since the continuation requires no further upstream input, those dynamics bring Bob’s qubit into the teleported state, as in Eq.~\eqref{eq:down}.
Thus, teleportation succeeds while the complete upstream collection~$S_{t_c}$ remains available as a record from which~$\psi$ is tomographically recoverable.

Now repeat this construction~$N$ times in series, using the output qubit of round~$j$ as the input qubit of round~$j+1$ and freshly initializing the ancillary systems in each round.
In the exact-copying limit, each round outputs a qubit in the state~$\ket{\psi}$.
By operational equivalence of pure states, the retained collections \(\{S_{t_c,j}(\psi)\}_{j=1}^N\) have the same joint operational statistics as \(N\) independent repetitions, although only one original unknown qubit was used.
For each \(N\), the copying accuracy can be chosen so that the recovery statistics of the entire \(N\)-round construction differ arbitrarily little from those of exact copying, with the discrepancy vanishing as \(N\to\infty\).
The recovery task
\[
\mathsf{Rec}_{S_{t_c}}^N:
\bigl(S_{t_c,1}(\psi),\ldots,S_{t_c,N}(\psi)\bigr)
\longrightarrow \widehat{\psi}_N
\]
yields~$d(\widehat{\psi}_N,\psi)\to0$ with arbitrarily high reliability.
From~$\widehat{\psi}_N$, arbitrarily many systems can then be prepared in states approaching~$\ket{\psi}$.
Because the construction uses only one original unknown state, this contradicts the no-cloning principle.
\hfill \(\square\)

% If the Einstein-local dynamics producing~$S_{t_c}(\psi)$
% admits a physical inverse, one can copy~$C$, complete
% teleportation using~$C'$, and apply the inverse dynamics
% to the retained~$S_{t_c}$, recovering Alice's input
% alongside Bob's output and violating no-cloning.
% %
% This parallels the copy-and-reverse argument of Galley, Giacomini, and Selby~\cite{Galley2023anyconsistent}, which rules out reversible information flow from a fully nonclassical system to a classical one.
% %
% Working within generalized probabilistic theories (GPTs), they define information flow operationally, through the dependence of one system's measurement statistics on another's preparation.
% %
% By this criterion, the two-bit classical message in teleportation carries no information about~$\psi$, since its outcome probabilities are independent of~$\psi$.
% %
% In the Deutsch--Hayden account, however, the physical systems carrying that message acquire a~$\psi$-dependence that measurements on those systems alone cannot reveal.
% %
% LMI thus captures a physical propagation of information that this operational criterion does not register.
% %
% Our theorem shows that, for teleportation, the obstruction persists even without reversible dynamics: under LMI, tomographic recoverability suffices.
% %
% In Appendix~\ref{app:robustness}, we derive quantitative fidelity
% bounds under additional operational and recovery assumptions.

The theorem requires neither reversible dynamics nor a specific quantum–classical interface, nor a particular model for the communication channel. 
Under LMI, teleportation through a medium thus becomes a test of its fundamental nature: perfect teleportation witnesses the medium's nonclassicality.
What happens if gravity itself carries the two bits?
%Gravity is the most consequential candidate.
%
The longstanding difficulty of reconciling quantum theory with gravity has motivated both gravity-related modifications of quantum dynamics~\cite{diosi1989models,penrose1996gravity} and approaches in which gravity or spacetime remains fundamentally classical~\cite{kafri2014,oppenheim2023prx,tilloydiosi2016}.
%
%The longstanding difficulty of reconciling quantum theory with gravity has motivated several approaches in which gravity or spacetime remains fundamentally classical~\cite{diosi1989models, penrose1996gravity, kafri2014, oppenheim2023prx, tilloydiosi2016}. 

\begin{figure}[t]
  \centering
  \includegraphics[width=\columnwidth, trim=9.6cm 7.75cm 7.7cm 11.1cm, clip]{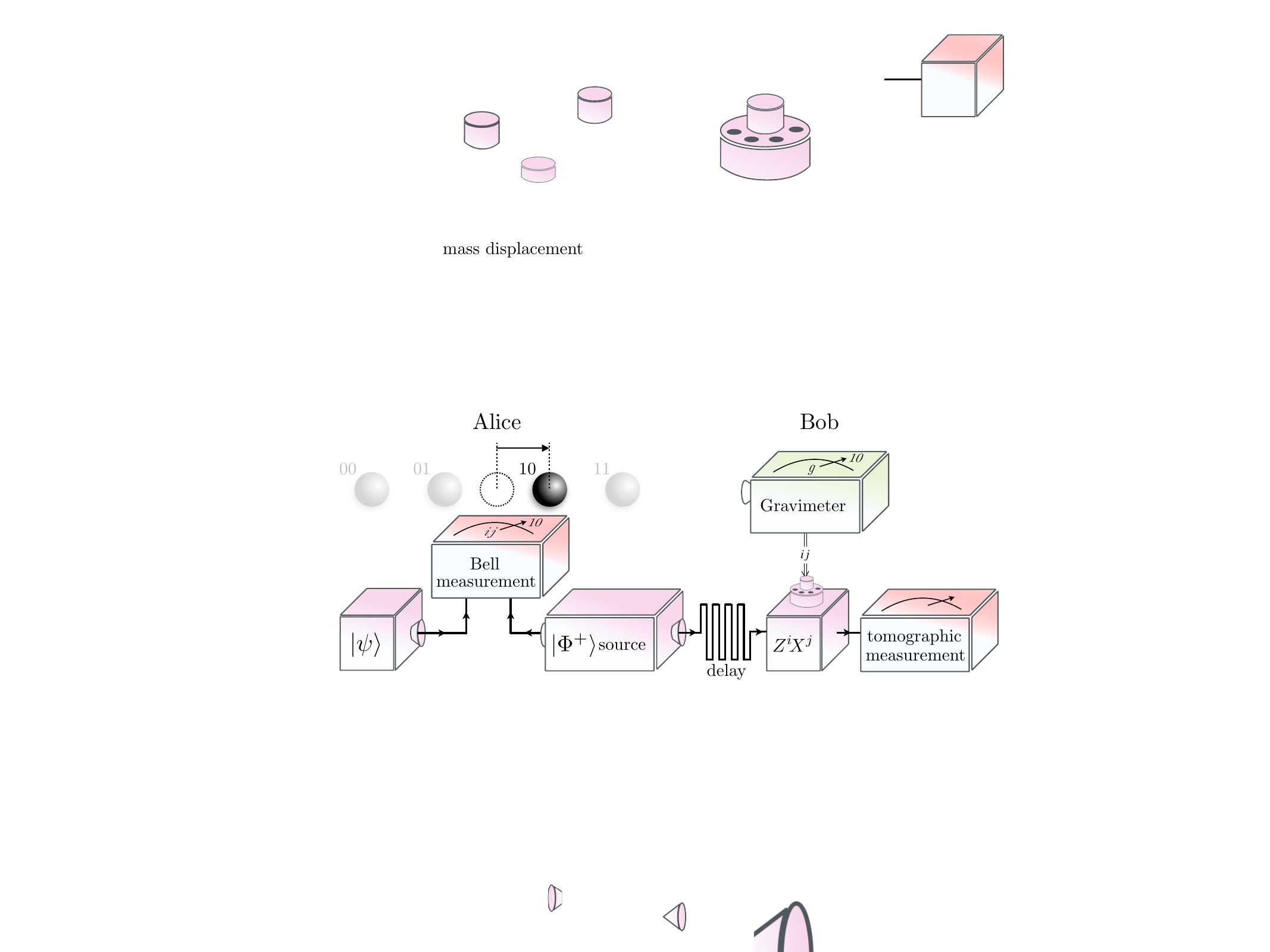}
  \caption{\textbf{Gravitational teleportation (GT).}
Alice's Bell-measurement outcome~$ij$ determines the
position of a source mass among four alternatives
(shown for~$ij=10$).
Bob infers~$ij$ from the gravitational acceleration~$g$
measured by his gravimeter.
}
  \label{fig:exp}
\end{figure}

\emph{We therefore propose a new experiment: Gravitational Teleportation~(GT).}
The protocol differs from standard quantum teleportation only in the communication step.
Alice encodes her measurement outcome~$(i,j)$ in one of four positions of a source mass.
%is encoded in a controlled mass displacement that faithfully represents the two-bit label~$(i,j)$.
%
Bob detects the resulting gravitational signal with a gravimeter and conditions his Pauli correction solely on that signal~(Fig.~\ref{fig:exp}).

Crucially, GT does not require coherence to be preserved in the source masses or the gravitational signal.
It instead probes the distinction between effective and fundamental classicality: under LMI, a decohered but fundamentally quantum channel can in principle support teleportation, whereas a fundamentally classical channel cannot, even if it faithfully transmits Alice's two-bit Bell label.
The theorem alone does not determine Bob's final state, but in Appendix~\SMref{sec:classical-bottleneck}{A} of the Supplemental
Material~\cite{supplemental} we illustrate how one possible failure mechanism, based on c-number-selected Pauli corrections, leaves his qubit maximally mixed.
Conversely, under LMI, perfect GT would establish that the gravitational mediator~\(C\) is not fundamentally classical.
%Conversely, under LMI, perfect GT would rule out a fundamentally classical gravitational medium, provided a collection~\(C\) of its degrees of freedom satisfies the continuation condition in Eq.~\eqref{eq:down} and non-gravitational communication pathways are excluded.
%
This would not establish that gravity obeys standard quantum dynamics, but only that its relevant degrees of freedom cannot be universally copied.
%the relevant gravitational degrees of freedom are not fundamentally classical---in the sense that their complete local descriptions cannot be universally copied.
%
Alternatively, if gravity---or any other communication medium---were independently established to be fundamentally classical, then perfect teleportation through that medium would instead refute LMI as a universal principle.

% Competing explanations
We can now return to the competing explanations of teleportation.
If teleportation is not locally mediated---as in accounts positing retrocausality, nonlocal state collapse, Bohmian guidance, global properties of the quantum state, or epistemic updating---the communication channel acts merely as a conduit for two discrete classical bits.
The carrier is essential for enforcing no-superluminal-signalling, but its fundamental degrees of freedom do not bear the \mbox{$\psi$-dependence}.
Consequently, these accounts give no reason for GT to fail whether the gravitational mediator is quantum or fundamentally classical.

By contrast, the Deutsch--Hayden account realizes LMI: the channel physically carries the~$\psi$-dependence through an operator-valued structure~\cite{deutsch2000information, bedard2023teleportation}.
If gravity is quantum, its degrees of freedom could play the same mediating role as decohered quantum systems in ordinary communication, so GT is expected to succeed.
The existing Deutsch--Hayden account has no fundamentally classical sector. Our theorem nevertheless applies to any LMI theory with a fundamentally classical gravitational mediator, including extensions of Deutsch--Hayden: perfect GT must fail.
Table~\ref{table:emp} summarizes these conditional expectations.
%The fundamentally classical case is different. The existing Deutsch–Hayden account contains no fundamentally classical sector, so this case lies beyond its present formulation. Nevertheless, our theorem constrains any extension of the Deutsch–Hayden mechanism—or any other LMI theory—with a fundamentally classical gravitational mediator: in every such theory, perfect GT must fail.  The resulting conditional structure is summarized in Table~\ref{table:emp}.

\begin{table}[t]
\centering
\small
\setlength{\tabcolsep}{3pt}
\renewcommand{\arraystretch}{1.25}
\begin{tabular}{@{}lcc@{}}
\hline\hline
& \makecell{Quantum gravity}
& \makecell{Classical gravity} \\
\hline
\makecell[l]{Non-LMI accounts}
& \makecell{expected to succeed}
& \makecell{expected to succeed} \\
\makecell[l]{LMI, DH-type acc.}
& \makecell{expected to succeed}
& \makecell{\textbf{GT must fail}} \\
\hline\hline
\end{tabular}
%
%\begin{table}[t]
%\centering
%\setlength{\tabcolsep}{3pt}
%\renewcommand{\arraystretch}{1.30}
%\begin{tabular}{@{}lcc@{}}
%\hline\hline
%& Quantum gravity & Classical gravity \\
%\hline
%{Non-LMI accounts~\cite{penrose1998quantum, jozsa2004illustrating, Zeilinger2007Teleportation, grw1986, pearle1989, bassi2003, bassi2013, maroney1999quantum, vaidman2022transfer, braunstein1996quantum, timpson2006grammar, fuchs2019qbism}}
%& GT expected to succeed & GT expected to succeed \\
%{LMI, DH-type mechanism~\cite{deutsch2000information, bedard2023teleportation}}
%& GT expected to succeed & \textbf{GT must fail} \\
%\hline\hline
%\end{tabular}
\caption{Conditional outcomes of Gravitational Teleportation (GT) under alternative accounts of teleportation and hypotheses about the gravitational mediator. The expectations of success are based on the available frameworks for those combinations of assumptions. The failure for classical gravity to locally mediate teleportation follows from our theorem.
}
\label{table:emp}
\end{table}

\paragraph{Relation to other work.}
Galley, Giacomini, and Selby~\cite{Galley2023anyconsistent} showed that reversible interactions cannot transfer information from a fully nonclassical system to a classical one. Working within generalized probabilistic theories, they define information flow operationally, through the dependence of one system’s measurement statistics on another’s preparation. By this criterion, the two-bit message in teleportation carries no information about \(\psi\). Yet the Deutsch–Hayden account traces the \(\psi\)-dependence through the physical carriers of those same bits, even though measurements on those carriers alone reveal no such dependence. The absence of locally observable information therefore need not imply the absence of physical mediation. LMI turns this distinction into a constraint on physical theories: together with no-cloning, it rules out perfect teleportation through a fundamentally classical channel.

\emph{Gravitationally induced entanglement} (GIE) proposals seek to witness the nonclassicality of gravity by testing whether two initially unentangled quantum probes can become entangled through an interaction mediated only by gravity~\cite{bose2017spin,marletto2017grav_ent}.
Two strands of reasoning underlie these proposals. 
Bose \emph{et al.} invoke the local-operations-and-classical-communication (LOCC) framework, which forbids entanglement generation from initially separable probes~\cite{bose2017spin}, while constructor-theoretic formulations of GIE make Einstein locality explicit~\cite{marletto_witnessing_2020,pietra_role_2025}, a condition also present in LMI.
GT instead begins with pre-shared entanglement and uses gravity to transmit the two-bit outcome.
Since non-LMI accounts need not assign the $\psi$-dependence to the gravitational channel, 
perfect GT may leave open a wider range of classical-gravity theories than a positive GIE result would. 
%successful GT may be less discriminatory among theories allowing classical gravity than successful GIE.
%
Notably, the LOCC framework itself, when applied to GT, predicts perfect teleportation using an abstract classical communication channel~\cite{bennett1993teleporting}.
The distinction between the LOCC strand and the LMI account developed here emerges when classical communication is examined as a physical process: while the former treats it as an abstract resource, the latter demands a complete local account of how the $\psi$-dependence reaches Bob.

Entanglement swapping brings this distinction back to the mediation of entanglement.
Suppose Alice teleports to Bob a qubit initially maximally entangled with Charlie's, using a second entangled pair shared with Bob and a gravitational feed-forward channel stipulated to be fundamentally classical.
The standard LOCC description predicts that, if the protocol is ideal, Bob and Charlie become maximally entangled, although they shared no entanglement initially~\cite{horodecki_quantum_2009}.
Under LMI and no-cloning, however, a fundamentally classical channel cannot implement this swap by a protocol that perfectly teleports arbitrary inputs.
%cannot support the perfect teleportation required for this swap.
%
Any complete physical account of entanglement mediation should therefore confront not only entanglement generation, but also entanglement swapping—and the mechanism of teleportation through which it can occur.

\paragraph{Feasibility.}

GT has a major experimental advantage over the GIE proposals discussed above and proposals based on gravitationally mediated state transfer~\cite{lami2024testing,toccacelo2025benchmarks}: it does not require coherence to be preserved in either the source masses or the gravitational degrees of freedom carrying the signal.
Alice's Bell label may instead be encoded and transmitted gravitationally through a macroscopically classical, decohered signal. After Alice’s Bell measurement, only Bob’s quantum memory must remain coherent until the feed-forward operation is completed.
Gravitational communication has been demonstrated at approximately one bit per minute~\cite{groszek2025}, while trapped-ion experiments have demonstrated quantum-state storage over several minutes~\cite{wang2021memory} and integrated remote entanglement with conditional quantum operations~\cite{main2025distributed}. These capabilities provide a plausible route to a proof-of-principle implementation using macroscopic source masses and a long-lived quantum memory. 
Their combination remains to be demonstrated and will require characterizing signalling errors and memory degradation over the full feed-forward interval.

Experimentally, applying the theorem requires identifying a mediator~\(C\) within the gravitational degrees of freedom and excluding non-gravitational pathways that could contribute to the transfer of the \(\psi\)-dependence.
Under LMI, any input-state dependence observed at Bob must then
have propagated through gravity.
This alone does not establish nonclassicality, since classical
communication can convey partial information about an unknown
quantum state~\cite{massar1995optimal}.
Turning our ideal no-go result into an experimental witness
therefore requires a quantitative bound on imperfect teleportation.
Appendix~\SMref{app:robustness}{B} of the Supplemental
Material~\cite{supplemental} derives fidelity bounds under additional operational assumptions and progressively stronger upstream recovery guarantees. These additional assumptions give the bounds a narrower scope than the main theorem.
%
%A concrete GT witness must therefore justify those assumptions, establish a bound valid across the class of theories being tested, and account for experimental imperfections.
%
Identifying weaker, physically motivated assumptions that yield
experimentally useful fidelity bounds remains an open problem.

\paragraph{Conclusion.}

Fundamentally classical systems cannot mediate perfect quantum teleportation under LMI and no-cloning. 
Assuming no-cloning, perfect gravitational teleportation would imply that either gravity has nonclassical degrees of freedom or LMI fails.

%\comc{Sentences to re-use?}
%The Deutsch--Hayden account turns the nature of the ``classical channel'' from an implementation detail into a substantive physical question.
%
%Classicality in teleportation is not an innocuous label, but a substantive claim about the underlying structure of physical systems.
%Gravitational teleportation thus becomes a diagnostic tool for probing the physical nature of gravity and, potentially, for adjudicating between competing explanatory frameworks of quantum mechanics.
%

\paragraph{Acknowledgements.}
The authors thank Xavier Coiteux-Roy, Claude Crépeau, David Deutsch and Lev Vaidman for their insightful comments, and the Conjecture Institute for its intellectual support, particularly Logan Chipkin for feedback on the manuscript.
CAB acknowledges support from a Mitacs Elevate postdoctoral fellowship in partnership with Bbox Digital. SR acknowledges support from the Israel Science Foundation Grant No. 1713/24.

\paragraph{AI Use Disclosure.}
We used ChatGPT (OpenAI; GPT-5.6 Sol and GPT-6 Astra) to help review the scientific arguments and check the literature and references. We also used GPT-6 Astra to help derive the fidelity bounds in Appendix~\SMref{app:robustness}{B} of the Supplemental Material~\cite{supplemental}---we directed and verified those derivations. We take full responsibility for the manuscript.

\SMreferences
\bibliography{references}

\BeginSupplement
\section{A classical bottleneck in the teleportation mechanism}
\label{sec:classical-bottleneck}

In the main text we established that, under locally mediated information (LMI) and the no-cloning principle, perfect teleportation through a mediator implies that it is not fundamentally classical.
The theorem does not determine Bob's final state.
Here we construct a partial hybrid quantum--classical model in which the channel faithfully transmits the Bell label, but Bob's qubit remains maximally mixed after the prescribed correction.
The calculation exhibits what changes when the operator-valued control of the Deutsch--Hayden mechanism is replaced by a realized classical state.
We specify only the local dynamics needed for this calculation, without claiming a complete hybrid theory.

\subsection{Local descriptions and the classical record}

Let us start by considering Alice's input qubit~$A_1$, her share~$A_2$ of the entangled pair, and Bob's qubit~$B$.
We take their complete local descriptions to be the Heisenberg-picture descriptors~\cite{deutsch2000information,bedard2021abc}
\begin{equation}
    \boldsymbol q_K=(q_{Kx},q_{Kz}),
    \qquad q_{Ky}=\mathrm{i}q_{Kx}q_{Kz},
    \qquad K\in\{A_1,A_2,B\}.
\end{equation}
The initial generators~$q_{Kw}^{(0)}$ act as the Pauli matrix~$\sigma_w$ on~$K$ and as the identity on the other qubits.
All quantum systems are initially in~$\ket{0}$, so the fixed Heisenberg reference state is~$\ket{\boldsymbol 0}$.
The preparation of~$\ket{\psi}$ is included in the evolution of~$A_1$'s descriptor.
We write~$\langle O\rangle=\bra{\boldsymbol 0}O\ket{\boldsymbol 0}$.

Entanglement preparation between~$A_2$ and~$B$ consists of a Hadamard gate followed by a CNOT.
Their action on the descriptors is as follows~\cite[\S~5]{bedard2021abc}:
\begin{center}
\begin{tikzpicture}[line width=0.8pt]
  \def\xA{0}
  \def\xB{4.8}
  \def\yH{1.45}
\def\yC{3.00}

  \tikzset{desc/.style={fill=white,inner sep=3pt,outer sep=0pt}}

  % Quantum wires, with a break at the Hadamard gate.
  \draw (\xA,0) -- (\xA,\yH-0.3);
  \draw (\xA,\yH+0.3) -- (\xA,4.3);
  \draw (\xB,0) -- (\xB,4.3);

  % Hadamard and CNOT.
  \draw (\xA-0.3,\yH-0.3) rectangle (\xA+0.3,\yH+0.3);
  \node at (\xA,\yH) {$H$};
  \fill (\xA,\yC) circle (2pt);
  \draw (\xA,\yC) -- (\xB,\yC);
  \draw (\xB,\yC) circle (0.22);
  \draw (\xB-0.22,\yC) -- (\xB+0.22,\yC);
  \draw (\xB,\yC-0.22) -- (\xB,\yC+0.22);

  % Descriptors at successive stages.
  \node[desc] at (\xA,0.65)
    {$\bigl(q_{A_2x}^{(0)},q_{A_2z}^{(0)}\bigr)$};
  \node[desc] at (\xB,0.65)
    {$\bigl(q_{Bx}^{(0)},q_{Bz}^{(0)}\bigr)$};
  \node[desc] at (\xA,2.25)
    {$\bigl(q_{A_2z}^{(0)},q_{A_2x}^{(0)}\bigr)$};
  \node[desc] at (\xA,3.75)
    {$\bigl(q_{A_2z}^{(0)}q_{Bx}^{(0)},q_{A_2x}^{(0)}\bigr)$};
  \node[desc] at (\xB,3.75)
    {$\bigl(q_{Bx}^{(0)},q_{A_2x}^{(0)}q_{Bz}^{(0)}\bigr)$};

  \node[above] at (\xA,4.3) {$A_2$};
  \node[above] at (\xB,4.3) {$B$};
  \draw[-{Triangle[length=1.5mm,width=1.5mm]}]
    (-2.4,0.4) -- (-2.4,1.5);
  \node[rotate=90,anchor=south] at (-2.45,0.95) {time};
\end{tikzpicture}
\end{center}
Thus, after entanglement preparation, Bob's descriptor is
\begin{equation}
    \boldsymbol q_B^{(1)}
    =\bigl(q_{Bx}^{(0)},q_{A_2x}^{(0)}q_{Bz}^{(0)}\bigr),
    \label{eq:sm-bell-descriptor}
\end{equation}
whose components satisfy
\begin{equation}
    \langle q_{Bx}^{(1)}\rangle
    =\langle q_{By}^{(1)}\rangle
    =\langle q_{Bz}^{(1)}\rangle=0.
    \label{eq:sm-zero-expectations}
\end{equation}

After preparing~$\ket{\psi}$, Alice performs the Bell-basis transformation: a CNOT from~$A_1$ to~$A_2$, followed by a Hadamard on~$A_1$.
We model the subsequent quantum--classical interface stochastically.
It produces a classical state~$\lambda\in\Lambda$ with distribution~$p_C(\lambda\mid\psi)$.
The realized~$\lambda$ is the complete local description of the mediator~$C$ and is copyable in the sense required in the main text.
The Bell label is a readout of this state,
\begin{equation}
    \ell:\Lambda\longrightarrow\{0,1\}^2,
    \qquad
    \ell(\lambda)=\bigl(i(\lambda),j(\lambda)\bigr),
\end{equation}
with the usual probabilities~$p(i,j\mid\psi)=1/4$.
In the simplest instantiation,~$\lambda=(i,j)$ and~$\ell$ is the identity map.
More generally,~$\Lambda$ may include additional discrete or continuous degrees of freedom, and~$\ell$ is a coarse-grained readout.
We leave~$p_C(\lambda\mid\psi)$ unspecified and allow it to depend on~$\psi$ beyond the uniformly distributed Bell label.

We impose Einstein locality: Alice's operations, including her coupling to~$C$, leave Bob's complete descriptor unchanged.
Before the feed-forward interaction, his descriptor is therefore~$\boldsymbol q_B^{(1)}$ in every realization.
We do not specify the backreaction on Alice's descriptors, as they play no further role in the downstream calculation.

\subsection{The classical-to-quantum coupling}

The realized mediator state~$\lambda$ selects Bob’s Pauli correction
\begin{equation}
    V(\lambda)=Z^{i(\lambda)}X^{j(\lambda)}.
\end{equation}
We assume that the classical-to-quantum coupling acts on his descriptor by the ordinary local-unitary rule, with the transformation selected by the realized classical state.
This is a substantive assumption about the coupling.

For the remainder of the calculation, write~$\boldsymbol q_B=\boldsymbol q_B^{(1)}$, suppressing the time index.
The functional representation of a gate $G$, which we denote~  $\widetilde G$, expresses it through sums and products of the descriptor components of the systems on which it acts.
Evaluated on the current (time-evolved) descriptors, this expression gives the unitary that updates them by conjugation~\cite[Sec.~4]{bedard2021abc}.
For the selected Pauli correction, this gives
%In terms of the current descriptors, the functional representation of the selected gate is
\begin{equation}
    \widetilde V(\lambda)
    =q_{Bz}^{\,i(\lambda)}q_{Bx}^{\,j(\lambda)},
    \qquad
    \boldsymbol q_B^{\mathrm{out}}(\lambda)
    =\widetilde V^\dagger(\lambda)
     \boldsymbol q_B\widetilde V(\lambda).
    \label{eq:sm-classical-control}
\end{equation}
Thus the four possible corrections act as follows:
\begin{center}
\renewcommand{\arraystretch}{1.25}
\setlength{\tabcolsep}{12pt}
\begin{tabular}{@{}ccc@{}}
\hline\hline
$\ell(\lambda)$ & Applied gate & Resulting descriptor on $B$ \\
\hline
$(0,0)$ & $\one$ & $\bigl(q_{Bx},q_{Bz}\bigr)$ \\
$(0,1)$ & $X$ & $\bigl(q_{Bx},-q_{Bz}\bigr)$ \\
$(1,0)$ & $Z$ & $\bigl(-q_{Bx},q_{Bz}\bigr)$ \\
$(1,1)$ & $ZX$ & $\bigl(-q_{Bx},-q_{Bz}\bigr)$ \\
\hline\hline
\end{tabular}
\end{center}
Each row applies to every classical state~$\lambda$ with the corresponding readout~$\ell(\lambda)$.

The observable statistics are captured by the reduced density matrix, which can be obtained from the descriptor components.
For a descriptor~$\boldsymbol r=(r_x,r_z)$ of~$B$, the reconstruction map is
\begin{equation}
    \Phi_B(\boldsymbol r)
    =\frac12\left(
       \one_B+\sum_{w\in\{x,y,z\}}
       \langle r_w\rangle\sigma_w
     \right),
    \qquad r_y=\mathrm{i}r_xr_z.
    \label{eq:sm-readout}
\end{equation}
Here the~$\sigma_w$ are Pauli matrices on Bob's single-qubit space.
We retain this fixed-reference readout throughout the hybrid model.
Applying it to Bob's post-correction descriptor, with~$\ell(\lambda)=(i,j)$, gives
\begin{equation}
\begin{aligned}
    \rho_B^{\mathrm{out}}(\lambda)
    &=\Phi_B\bigl(\boldsymbol q_B^{\mathrm{out}}(\lambda)\bigr)\\
    &=\frac12\Bigl[
      \one_B
      +(-1)^i\langle q_{Bx}\rangle\sigma_x
      +(-1)^{i+j}\langle q_{By}\rangle\sigma_y
      +(-1)^j\langle q_{Bz}\rangle\sigma_z
      \Bigr]
      =\frac{\one_B}{2},
\end{aligned}
\label{eq:sm-mixed-output}
\end{equation}
as all three expectation values vanish by Eq.~\eqref{eq:sm-zero-expectations}.
Bob's qubit therefore remains maximally mixed for every realized classical state, and hence also on average:
\begin{equation}
    \overline{\rho}_B^{\mathrm{out}}(\psi)
    =\int_{\Lambda}\!d\lambda\,
      p_C(\lambda\mid\psi)\rho_B^{\mathrm{out}}(\lambda)
    =\frac{\one_B}{2}.
\end{equation}

%%
%Bob's qubit therefore remains maximally mixed for every realized classical state, not merely on average.
%%
%%For every pure input, the fidelity is
%%\begin{equation}
%%    F(\psi)=\bra{\psi}\rho_B^{\mathrm{out}}(\lambda)\ket{\psi}
%%    =\frac12.
%%\end{equation}
%%\comc{Not sure it's worth keeping the fidelity calculation here.}
%Averaging over realizations of the stochastic interface gives the same result:
%\begin{equation}
%    \rho_B^{\mathrm{out}}(\psi)
%    =\int_{\Lambda}\!d\lambda\,
%      p_C(\lambda\mid\psi)\rho_B^{\mathrm{out}}(\lambda)
%    =\frac{\one_B}{2}.
%\end{equation}
%%For discrete~$\Lambda$, the integral is replaced by a sum.
%%

The conclusion does not depend on the detailed form of~$p_C(\lambda\mid\psi)$.
Nor is the failure due to restricting the quantum correction to the four Pauli corrections. Even if the classical state selects an arbitrary single-qubit unitary, its action only mixes the current descriptor components, whose expectation values all vanish. Bob’s output therefore remains maximally mixed.
An additional stochastic choice among such unitaries at the classical-to-quantum interface would therefore also leave the result unchanged.

\subsection{The difference from quantum control}

Why does the same calculation not rule out ordinary LMI, DH-like quantum teleportation?
The crucial difference is not the list of Pauli corrections, but the physical interaction that controls them.
In the fully quantum analysis, the systems implementing the records remain quantum, even when decohered, and their descriptors supply operator-valued control.

Let~$\Pi_{ij}(\psi)$ denote the four orthogonal record projectors immediately before the feed-forward interaction.
They are constructed from the current descriptors of the (quantum) record systems, satisfy~$\sum_{i,j}\Pi_{ij}(\psi)=\one$, and commute with Bob's descriptor components.
The reference state is extended to include the initially blank quantum records.
The quantum feed-forward unitary has the functional representation
\begin{equation}
    \widetilde U_{\mathrm{ff}}
    =\sum_{i,j=0}^1
      \Pi_{ij}(\psi)\,q_{Bz}^{\,i}q_{Bx}^{\,j}.
    \label{eq:sm-quantum-control}
\end{equation}
Conjugating Bob's descriptor by this unitary yields
\begin{equation}
\begin{aligned}
    q_{Bx}^{\mathrm{out}}
    &=\sum_{i,j=0}^1
      (-1)^i\Pi_{ij}(\psi)q_{Bx},\\
    q_{Bz}^{\mathrm{out}}
    &=\sum_{i,j=0}^1
      (-1)^j\Pi_{ij}(\psi)q_{Bz}.
\end{aligned}
\label{eq:sm-quantum-output}
\end{equation}
Unlike the numerical signs in the classical model, these control operators carry~$\psi$-dependence into Bob's descriptor.
Although~$\langle q_{Bw}\rangle=0$, the expectations~$\langle\Pi_{ij}(\psi)q_{Bw}\rangle$ need not vanish.
In the usual protocol, the resulting sums recover the input-state expectation values, yielding~$\ket{\psi}\bra{\psi}$ on~$B$~\cite{deutsch2000information,bedard2023teleportation}.

A \emph{c-number-selected} Pauli correction is therefore not the same physical interaction as a \emph{q-number-selected} one.
The projectors in Eq.~\eqref{eq:sm-quantum-control} cannot be replaced by their probabilities, or by a sampled classical label, without changing the mechanism.
Decoherence in the record basis does not make this replacement: the relevant descriptor structure survives and can propagate through a chain of local interactions~\cite{bedard2023teleportation}.
In both analyses, Alice's operations leave Bob's descriptor unchanged.
What differs is the local interaction by which her physical record subsequently acts on it.

\subsection{Alternative couplings and the no-cloning constraint}

The coupling in Eq.~\eqref{eq:sm-classical-control} is an assumption, not a consequence of fundamental classicality alone.
One could instead postulate a different local rule,
\begin{equation}
    (\lambda,\boldsymbol q_B)
    \longmapsto\mathcal G_B(\lambda,\boldsymbol q_B),
\end{equation}
where only Bob's output descriptor is displayed.
Such a rule would have to specify both how the classical state encodes~$\psi$ and how the interaction brings that dependence into Bob's descriptor.
It cannot invoke~$\psi$ as an additional argument that has no local physical carrier.

Changing the distribution~$p_C(\lambda\mid\psi)$, or enriching the classical state space, does not by itself avoid the preceding result.
Even if the numerical parameters selecting a single-qubit unitary specify~$\psi$, conjugating Bob's current descriptor by a c-number selected single-qubit unitary then applying~$\Phi_B$ still gives~$\one_B/2$.
Thus recovering~$\rho_B^{\text{out}} = \ketbra{\psi}{\psi}$ would require a different downstream interaction~$\mathcal G_B$, not merely a more elaborate c-number selection of an ordinary unitary.

A directly readable encoding of~$\psi$ faces a further obstruction.
For example, suppose~$\lambda=(i,j,\chi)$ and the quantum--classical interface produces
\begin{equation}
    p_C(i,j,\chi\mid\psi)
    =\frac14\,\delta(\chi-\psi),
\end{equation}
where the Dirac delta denotes a distribution concentrated at~$\chi=\psi$.
Every realization would then contain a classical specification of the input state in addition to the Bell label.
If~$\chi$ can be read to arbitrary accuracy and used to prepare qubits in the corresponding state, a single unknown input would suffice to produce arbitrarily many arbitrarily faithful copies.
This contradicts the no-cloning principle~\cite{wootters1982,dieks1982,keyl1999}.

This argument does not prohibit~$p_C(\lambda\mid\psi)$ from depending on~$\psi$.
Such dependence need not make~$\psi$ faithfully readable from one realized classical state, and copying that realization does not supply independent samples from the distribution.
The theorem in the main text constrains perfect teleportation under LMI and no-cloning without assuming that~$\psi$ is directly readable from the channel.

We have not established that the partial dynamics specified here admits a complete extension satisfying LMI, including tomographic recoverability.
The fidelity of~$1/2$ is a consequence of this particular coupling, not a universal bound derived from the theorem.
The model isolates a definite failure mode: a classical instruction can select the prescribed Pauli gate without reproducing the operator-valued interaction through which LMI quantum teleportation transfers the state dependence.

%%% V3 AB

\section{Fidelity bounds under additional operational assumptions}
\label{app:robustness}

The main theorem establishes an exact obstruction: under LMI and the no-cloning principle, perfect teleportation through a fundamentally classical medium
is impossible.
Yet no finite experiment can establish fidelity-one teleportation, so an experimental witness must also constrain imperfect teleportation.
How close to fidelity one could teleportation come through such a
medium?
The answer depends on how LMI’s recoverability requirement is quantified or strengthened for the systems retained upstream, even as Bob receives the output qubit.
Using methods from quantum information theory under additional operational assumptions, we derive three bounds, corresponding to progressively stronger forms of recoverability: eventual tomography yields a fidelity bound strictly below one; a readout that reproduces the outcome probabilities of a full-strength measurement of the input qubit yields an upper bound of~$2/3$;
and exact quantum recovery of the input qubit yields an upper bound of~$1/2$.

These operational assumptions go beyond those of the main theorem and are specified separately for each bound.
The argument for the first bound assumes a quantum channel producing a joint density operator for the retained collection and Bob’s qubit; the other two arguments require only a quantum instrument or a two-qubit channel for the complete experiment, including recovery.
We first revisit the copying construction of the main proof, which retains the upstream systems while allowing teleportation to continue, and define the common fidelity benchmark.
We then derive the three bounds in turn, and finally discuss copying accuracy and the scope of the resulting witnesses.

\subsection{The copy-and-retain construction}
\label{app:robustness_setup}

Let \(R\equiv S_{t_c}\) comprise the entire upstream collection at \(t_c\), including the mediator \(C\).
%Let $R\equiv S_{t_c}$ comprise the entire upstream collection at the cut, including the downstream-sufficient section~$C$.
If $C$ is fundamentally classical, its complete local description can be copied onto an initially blank system~$C'$ without altering the original.
We retain $R$ and use \(C'\) in the downstream continuation.
We call this the \emph{copy-and-retain construction}.
Because the downstream continuation requires no further upstream input, Einstein locality ensures that Bob receives the same output as in the original protocol, whether teleportation is perfect or imperfect.
The retained collection remains available for recovery without further interaction with Bob.
We initially treat copying as exact and return to finite accuracy at the end.

The operational descriptions used below concern the joint predictions of the complete experiment, including the hybrid interface and the relevant recovery procedure, without specifying the mediator's microscopic dynamics.
Fundamental classicality still means universal copyability of~$C$'s complete local description, not merely of an accessible classical record.
These additional descriptions do not follow merely from assigning quantum states to Alice’s and Bob’s qubits.
The state assignments specify individual qubit measurement statistics.
The additional assumptions constrain the joint process producing Bob’s output and the retained systems or recovery output, requiring the corresponding maps to be completely positive.

All ancillary systems---the shared entangled pair, physical communication channel, apparatus, and environment---have a fixed joint initial preparation independent of the input state.
Recovery may act on all of $R$, but receives no independent information about the unknown input.

Writing $\rho_\psi=\ket{\psi}\bra{\psi}$, denoting Bob’s output qubit by~$B$ and its state by~$\rho_B^{\mathrm{out}}(\psi)$, and taking $d\psi$ to be normalized Haar measure, define
\begin{equation}
    F_{\mathrm{tel}}
    =\int d\psi\,\bra{\psi}\rho_B^{\mathrm{out}}(\psi)\ket{\psi}.
    \label{eq:rob_fidelity}
\end{equation}
Thus all pure input qubits are weighted uniformly over the Bloch sphere.
All bounds concern protocols without postselection, with every measurement outcome included in the average according to its probability.
They do not apply to fidelities conditioned on postselected successful trials.

\subsection{Eventual tomography}
\label{app:robustness_finiteN}

Eventual tomography demands the least recovery capability of the three cases considered here, but our proof of its fidelity bound uses the most extensive operational description.

For this argument, we assume that the copy-and-retain experiment admits a joint density-operator description of Bob's output qubit and the entire retained collection.
Incorporating the fixed ancillary preparation into the dynamics, one execution is represented by a completely positive, trace-preserving (CPTP) map
\begin{equation}
    \mathcal V:\rho\longmapsto\omega_{BR}(\rho),
    \qquad
    \mathcal T=\tr_R\circ\mathcal V,
    \qquad
    \mathcal E=\tr_B\circ\mathcal V.
    \label{eq:rob_channel}
\end{equation}
Here $\mathcal T$ gives Bob's output and $\mathcal E$ the retained collection.
These are the two marginals of the same physical process, so constraints on their simultaneous performance apply.
In particular, $\mathcal T(\rho_\psi)=\rho_B^{\mathrm{out}}(\psi)$.
Independent repetitions use fresh ancillary systems in the same preparation and the same dynamics.
We assume tensor-product composition for these repetitions and a quantum-measurement representation of the allowed recovery readout.
The density-operator representation may include classical degrees of freedom as block-diagonal sectors.
The next two arguments do not require this representation of the retained collection.

To quantify recovery, consider $N$ independently prepared retained collections, each generated from a fresh input qubit in the same unknown state~$\rho_\psi$.
No teleportation output is reused as a subsequent input.
An allowed measurement $\{M_x^{(N)}\}$ on these collections assigns a pure-state estimate $\ket{\widehat\psi_{N,x}}$ to outcome~$x$.
Its mean estimation fidelity is
\begin{equation}
    G_N=\int d\psi\sum_x
    \tr\!\left[M_x^{(N)}\mathcal E(\rho_\psi)^{\otimes N}\right]
    \left|\langle\widehat\psi_{N,x}|\psi\rangle\right|^2.
    \label{eq:rob_GN}
\end{equation}
The measurement may be collective, but neither it nor the assigned estimates may depend on~$\psi$.
This is a benchmark for recovery from independent preparations; $F_{\mathrm{tel}}$ still concerns one execution of teleportation.

We will show that any such benchmark satisfies
\begin{equation}
    \boxed{
    F_{\mathrm{tel}}\leq
    \frac23+\frac13
       \sqrt{1-\left(\frac{2G_N-1}{N}\right)^2},
    \qquad G_N\geq\frac12.}
    \label{eq:rob_finiteN_bound}
\end{equation}
An estimate independent of the input has Haar-averaged fidelity~$1/2$.
Any recovery advantage above this baseline therefore gives a strictly positive gap below teleportation fidelity one.
Tomographic recoverability requires a sequence of allowed recovery
procedures, independent of the unknown input, whose estimates become arbitrarily accurate with arbitrarily high probability for every input state as the number of independent preparations increases.
Consequently, the fidelity~$|\langle\widehat\psi_{N,x}|\psi\rangle|^2$ approaches one with probability approaching one.
Since this fidelity lies between zero and one, the increasingly rare
inaccurate estimates contribute vanishingly little to its mean.
Averaging over measurement outcomes and uniformly over input states
therefore gives $G_N\to1$.
Thus some finite~$N$ already satisfies~$G_N>1/2$, yielding a bound on teleportation fidelity strictly below one.
Tomographic recoverability alone specifies neither a convergence rate
nor an explicit finite~$N$ for which $G_N>1/2$ is guaranteed.
It therefore establishes the existence of a positive gap, while
evaluating its size requires a quantitative recovery guarantee.

\paragraph{From recovery to retained distinguishability.}
The first step is to translate the recovery benchmark into a statement about a single retained collection.
For trace distance $D(\rho,\sigma)=\tfrac12\|\rho-\sigma\|_1$, define
\begin{equation}
    \delta=\max_{|\mathbf n|=1}
      D\!\left(\mathcal E(\rho_{\mathbf n}),
               \mathcal E(\rho_{-\mathbf n})\right),
    \qquad
    \rho_{\mathbf n}=\tfrac12
       (\one+\mathbf n\cdot\boldsymbol\sigma).
    \label{eq:rob_delta}
\end{equation}
This quantity measures how well the retained output can distinguish some pair of orthogonal inputs.
For two equally likely retained states attaining
~\(\delta\), optimizing over all two-outcome quantum measurements (POVMs) on the retained collection~\(R\) gives a discrimination success probability of~\((1+\delta)/2\). Every trial must produce a guess of which state was prepared; inconclusive answers are excluded. The optimization includes joint measurements on the constituent systems of~\(R\).
We will use this mathematical optimum below to bound Bob’s fidelity; the argument does not require the optimizing measurement to be physically available in the hybrid theory.

Compare the retained states with the input-independent reference $\tau_0=\mathcal E(\one/2)$.
Since $\one/2=(\rho_{\mathbf n}+\rho_{-\mathbf n})/2$, linearity gives $D(\mathcal E(\rho_\psi),\tau_0)\leq\delta/2$ for every pure input.
Replacing the factors of a tensor product one at a time and applying the triangle inequality then gives
\begin{equation}
    D\!\left(\mathcal E(\rho_\psi)^{\otimes N},\tau_0^{\otimes N}\right)
    \leq N D\!\left(\mathcal E(\rho_\psi),\tau_0\right)
    \leq\frac{N\delta}{2}.
    \label{eq:rob_benchmark_distance}
\end{equation}
Apply the estimator used to define $G_N$ in Eq.~\eqref{eq:rob_GN}
to $\tau_0^{\otimes N}$, keeping its measurement and the guess
assigned to each outcome unchanged.
Its outcome probabilities are independent of~$\psi$, and each fixed qubit guess has Haar-averaged fidelity~$1/2$, so the estimator's
average fidelity is also~$1/2$.
For each~$\psi$, the fidelity score lies between zero and one,
so replacing the measured state changes its expected value by
at most the trace distance.
Using Eq.~\eqref{eq:rob_benchmark_distance} and averaging over~$\psi$
therefore gives
\begin{equation}
    G_N-\frac12\leq\frac{N\delta}{2}.
    \label{eq:rob_benchmark_advantage}
\end{equation}
Thus a recovery advantage at finite~$N$ requires input dependence
already in the state of a single retained collection, quantified by
$\delta\geq(2G_N-1)/N$.

\paragraph{From retained distinguishability to Bob's fidelity.}
Banaszek considers quantum operations without postselection that produce both an output quantum state and a classical outcome from which the input state is estimated~\cite{banaszek2001fidelity}.
Such an operation is called a quantum instrument.
He bounds the simultaneous quality of its two outputs, averaging both fidelities uniformly over pure input states.
For a qubit, the output fidelity $F$ and estimation fidelity $G$ obey
\begin{equation}
    F\leq\frac23+
       2\sqrt{\left(G-\frac13\right)
               \left(\frac23-G\right)},
    \qquad \frac12\leq G\leq\frac23.
    \label{eq:rob_banaszek}
\end{equation}
The bound also applies to nonoptimal guesses in this interval: optimizing them can only increase $G$, which decreases the right-hand side.

Our copy-and-retain construction has this structure once a readout is applied to~$R$: Bob's qubit is the quantum output, and the readout outcome supplies the estimate.
Ignoring that outcome leaves Bob's marginal channel unchanged, so the output fidelity in Eq.~\eqref{eq:rob_banaszek} is precisely~$F_{\mathrm{tel}}$.
We now construct a readout whose estimation fidelity is fixed by~$\delta$.

Choose a direction $\mathbf n$ attaining the maximum in Eq.~\eqref{eq:rob_delta}, and a two-outcome measurement $\{H,\one-H\}$ optimally distinguishing the corresponding retained states.
Linearity allows its outcome probability to be written as
$p(+|\mathbf r)=\tr[H\mathcal E(\rho_{\mathbf r})]=a+\mathbf b\cdot\mathbf r$.
Optimal discrimination gives $p(+|\mathbf n)-p(+|-\mathbf n)=\delta$, so $2\mathbf b\cdot\mathbf n=\delta$.
Assign the guesses $\rho_{\mathbf n}$ and $\rho_{-\mathbf n}$ to the two outcomes.
Their fidelities with an input of Bloch vector $\mathbf r$ are $(1\pm\mathbf n\cdot\mathbf r)/2$.
The mean estimation fidelity of this two-outcome estimator, averaged uniformly
over pure input qubits, is
\begin{equation}
\begin{aligned}
    G_{\mathrm{bin}}
    &=\int d\mathbf r\,\Biggl[
       p(+|\mathbf r)\frac{1+\mathbf n\cdot\mathbf r}{2}
       +\bigl(1-p(+|\mathbf r)\bigr)
        \frac{1-\mathbf n\cdot\mathbf r}{2}
       \Biggr]\\
    &=\frac12+\frac{\mathbf b\cdot\mathbf n}{3} \\
    &=\frac12+\frac{\delta}{6}.
\end{aligned}
\label{eq:rob_binary_estimation}
\end{equation}
To evaluate the integral, we used
$\int d\mathbf r\,r_i=0$ and
$\int d\mathbf r\,r_i r_j=\delta_{ij}/3$;
and the final equality follows from $2\mathbf b\cdot\mathbf n=\delta$.
Composing this readout mathematically with $\mathcal V$ defines a
quantum instrument with fidelities $F=F_{\mathrm{tel}}$ and
$G=G_{\mathrm{bin}}$.
Equation~\eqref{eq:rob_banaszek} therefore yields
\begin{equation}
    F_{\mathrm{tel}}\leq f(\delta),
    \qquad
    f(x)=\frac23+\frac13\sqrt{1-x^2},
    \quad 0\leq x\leq1.
    \label{eq:rob_distinguishability_bound}
\end{equation}
Since $f$ decreases, Eq.~\eqref{eq:rob_benchmark_advantage} gives
$F_{\mathrm{tel}}\leq f((2G_N-1)/N)$, proving Eq.~\eqref{eq:rob_finiteN_bound}.

The auxiliary readout need not be physically available in the
hybrid theory.
The assumed joint quantum-channel structure ensures that the
constructed instrument is mathematically valid.
If $\mathcal V$ violated Eq.~\eqref{eq:rob_distinguishability_bound},
this instrument would violate Banaszek's inequality,
Eq.~\eqref{eq:rob_banaszek}, which every such instrument must satisfy.

The proof constrains a single copy-and-retain execution; independent
repetitions enter only through the recovery benchmark.
A numerical witness valid across a class of models follows if
$G_N\geq g>1/2$ holds for every model at a common finite~$N$
and with a common lower bound~$g$.
Qualitative tomographic recoverability, as defined in the main text,
guarantees an advantage for each model at some finite~$N$, but does
not specify how large~$N$ must be.
It therefore establishes a positive fidelity gap for each model
separately without providing a common numerical gap across models;
the certified gap may be arbitrarily small as the model varies.

\subsection{Retaining a full-strength input measurement}
\label{app:robustness_measurement}

A stronger recoverability assumption is that an allowed readout of one retained collection reproduces the probabilities of a rank-one informationally complete POVM $\{E_y\}$ on the original input:
\begin{equation}
    p_R(y|\psi)=\tr(E_y\rho_\psi),
    \qquad
    E_y=w_y\ket{\chi_y}\bra{\chi_y},
    \qquad \sum_y E_y=\one.
    \label{eq:rob_measurement_reproduction}
\end{equation}
Informational completeness means that these probabilities determine the input density operator.
For example, the tetrahedral POVM has effects
$E_a=\tfrac14(\one+\mathbf n_a\cdot\boldsymbol\sigma)$,
where the four unit vectors point towards the vertices of a regular tetrahedron.
The hypothesis requires an allowed readout of one retained collection to reproduce the specified rank-one POVM’s outcome probabilities exactly, for every input. Informational completeness alone would also allow probabilities whose dependence on the input is arbitrarily weak.

We assume that the complete copy-and-retain experiment, including the readout of~$R$ after Bob's continuation, defines a quantum instrument
$\{\mathcal I_y\}$ from the input qubit to Bob's output qubit.
Such an instrument would follow from the joint CPTP description in Eq.~\eqref{eq:rob_channel}, together with a quantum-measurement representation of this readout.
Here we require only the instrument description, without a separate density-operator description of the retained collection.
Its branches are completely positive and trace nonincreasing, and their sum is trace preserving.
Including every readout outcome leaves Bob's output unchanged:
\begin{equation}
    \sum_y\mathcal I_y(\rho_\psi)=\rho_B^{\mathrm{out}}(\psi).
    \label{eq:rob_instrument_marginal}
\end{equation}
The branch $\mathcal I_y(\rho)$ is Bob's unnormalized state
associated with readout outcome~$y$.
Writing
$\mathcal I_y(\rho)=\sum_\mu K_{y\mu}\rho K_{y\mu}^\dagger$,
the measurement-reproduction hypothesis gives
\begin{equation}
    \sum_\mu K_{y\mu}^\dagger K_{y\mu}
    =E_y=w_y\ket{\chi_y}\bra{\chi_y}.
    \label{eq:rob_rank_one_effect}
\end{equation}
For any input vector $\ket{\phi}$ orthogonal to $\ket{\chi_y}$,
\begin{equation}
    0=\bra{\phi}E_y\ket{\phi}
     =\sum_\mu\|K_{y\mu}\ket{\phi}\|^2.
    \label{eq:rob_kraus_orthogonal}
\end{equation}
Since every term is nonnegative, each $K_{y\mu}$ annihilates
this orthogonal subspace.
Hence $K_{y\mu}=\ket{v_{y\mu}}\bra{\chi_y}$, where
$\ket{v_{y\mu}}=K_{y\mu}\ket{\chi_y}$, and therefore
\begin{equation}
\begin{aligned}
    \mathcal I_y(\rho)
    &=\bra{\chi_y}\rho\ket{\chi_y}
      \sum_\mu\ket{v_{y\mu}}\bra{v_{y\mu}}\\
    &=\tr(E_y\rho)\sigma_y.
\end{aligned}
\label{eq:rob_measure_prepare}
\end{equation}

Here
$\sigma_y=w_y^{-1}\sum_\mu\ket{v_{y\mu}}\bra{v_{y\mu}}$
is independent of~$\rho$ and has unit trace because
$\sum_\mu\langle v_{y\mu}|v_{y\mu}\rangle=w_y$.
Thus, conditional on~$y$, Bob's normalized state is fixed;
all input dependence lies in the outcome probabilities.
Summing over outcomes gives
$\sum_y\mathcal I_y(\rho)=\sum_y\tr(E_y\rho)\sigma_y$,
so Bob's channel is measure-and-prepare.

Massar and Popescu showed that the optimal mean fidelity for estimating a uniformly distributed pure qubit from one copy is $2/3$~\cite{massar1995optimal}.
A measure-and-prepare channel cannot exceed the same fidelity: its prepared state is determined solely by the classical measurement outcome.
Thus
\begin{equation}
    \boxed{F_{\mathrm{tel}}\leq\frac23.}
    \label{eq:rob_two_thirds}
\end{equation}
The rank-one condition enforces this bound; informational completeness additionally ensures that repeated readouts permit tomography.
The pre-shared entangled resource remains included in the fixed ancillary preparation.

When exact reproduction of the measurement statistics is imperfect, any established lower bound on the single-run estimation fidelity can be inserted directly into Eq.~\(\eqref{eq:rob_banaszek}\) to obtain the corresponding teleportation bound.

\subsection{Recovering the quantum input}
\label{app:robustness_quantum_recovery}

The strongest recovery considered here returns a qubit carrying the input state, rather than a classical estimate.
Suppose an allowed decoder on $R$ produces a qubit $A'$.
The decoder is independent of~$\psi$ and uses no downstream systems.
We assume that the complete copy-and-retain experiment, including this recovery after Bob's continuation, defines a completely positive, trace-preserving map from the input qubit to the joint output~$BA'$:
\begin{equation}
\begin{aligned}
    \mathcal W:\rho&\longmapsto\omega_{BA'}(\rho),\\
    \tr_{A'}\mathcal W(\rho_\psi)&=\rho_B^{\mathrm{out}}(\psi).
\end{aligned}
    \label{eq:rob_decoded_channel}
\end{equation}
The second equality expresses that recovery leaves Bob's output unchanged.
Only this joint input-to-output channel is assumed; no separate density-operator description of~$R$ or quantum-channel representation of the decoder is required.
Suppose the recovery fidelity satisfies
\begin{equation}
    F_{\mathrm{rec}}
    =\int d\psi\,
      \bra{\psi}\tr_B[\mathcal W(\rho_\psi)]\ket{\psi}
    \geq1-\eta.
    \label{eq:rob_recovery_fidelity}
\end{equation}
The joint channel produces two approximate copies from one unknown input, with fidelities $F_{\mathrm{tel}}$ and $F_{\mathrm{rec}}$.
The copies need not have equal quality, so this is an asymmetric $1\!\to\!2$ cloning task.

The optimal asymmetric-cloning results of Cerf~\cite{cerf2000pauli} and of Fiur\'a\v{s}ek, Filip, and Cerf~\cite{fiurasek2005asymmetric} constrain the two Haar-averaged output fidelities of any quantum channel from
one qubit to two, without postselection.
The fidelity of either output may vary with the input state;
the bounds concern its uniform average over pure inputs.
They therefore apply directly to the joint channel in
Eq.~\eqref{eq:rob_decoded_channel}.
Writing $e_A=1-F_{\mathrm{rec}}$ and $e_B=1-F_{\mathrm{tel}}$, the qubit trade-off is
\begin{equation}
    e_A+e_B+\sqrt{e_Ae_B}\geq\frac12.
    \label{eq:rob_cloning_tradeoff}
\end{equation}
Using $e_A\leq\eta$ and solving for $e_B$ yields
\begin{equation}
    F_{\mathrm{tel}}
    \leq\frac12+\frac\eta2+
       \frac12\sqrt{\eta(2-3\eta)},
    \qquad 0\leq\eta\leq\frac12.
    \label{eq:rob_approx_quantum_recovery}
\end{equation}
For example, $F_{\mathrm{rec}}\geq0.99$ implies $F_{\mathrm{tel}}\leq0.5752$ approximately.

Setting $\eta=0$ in Eq.~\eqref{eq:rob_approx_quantum_recovery}
gives
\begin{equation}
    \boxed{
    F_{\mathrm{tel}}\leq\frac12
    \qquad \text{if } F_{\mathrm{rec}}=1.}
    \label{eq:rob_one_half}
\end{equation}

In fact, exact recovery forces Bob's output to be independent
of the input.
To see this, note that a Haar-averaged recovery fidelity of one
forces the continuous, nonnegative statewise infidelity to
vanish for every pure input, so $\tr_B\circ\mathcal W$
is the identity channel.
The no-information-without-disturbance result of Kretschmann,
Schlingemann, and Werner implies that a joint quantum channel
with an identity marginal must have a constant other
marginal~\cite{kretschmann2008continuity}.
Applied to Eq.~\eqref{eq:rob_decoded_channel}, this means that
Bob's output is a fixed state $\sigma_B$, independent of~$\psi$.
Its Haar-averaged
fidelity is
\begin{equation}
    F_{\mathrm{tel}}
    =\tr\!\left(\sigma_B\int d\psi\,\rho_\psi\right)
    =\tr(\sigma_B\one/2)
    =\frac12.
    \label{eq:rob_constant_fidelity}
\end{equation}
The output need not be maximally mixed: every fixed normalized
qubit state has this Haar-averaged overlap.

If the evolution from input preparation to the cut admits an
allowed inverse acting on $R$ alone, exact input recovery follows,
as in a copy-and-reverse argument~\cite{Galley2023anyconsistent}.
Full reversal is sufficient but not necessary: the decoder need
only recover the input qubit, without restoring the apparatus
to its initial state.

The recovery capabilities form a hierarchy: exact quantum recovery permits any input measurement, and reproduction of an informationally complete input measurement permits eventual tomography.
Their quantitative consequences differ because eventual recovery from many preparations need not imply strong recovery from one.
The complete assumption sets are not thereby ordered, because the operational descriptions differ.

\begin{table}[htbp]
\centering
\caption{Recovery guarantees and operational descriptions assumed for the three bounds.}
\label{tab:rob_assumptions}
\small
\setlength{\tabcolsep}{4pt}
\renewcommand{\arraystretch}{1.15}

\begin{tabular}{@{}lll@{}}
\hline\hline

\parbox[t]{0.15\linewidth}{\raggedright
Bound}
&
\parbox[t]{0.34\linewidth}{\raggedright
Recovery guarantee}
&
\parbox[t]{\dimexpr0.51\linewidth-4\tabcolsep\relax}{\raggedright
Operational description assumed for that argument}
\\
\hline

\parbox[t]{0.15\linewidth}{\raggedright
Eventual tomography}
&
\parbox[t]{0.34\linewidth}{\raggedright
An allowed estimator achieves $G_N>1/2$ at some finite $N$;
eventual tomography guarantees that such a benchmark exists.}
&
\parbox[t]{\dimexpr0.51\linewidth-4\tabcolsep\relax}{\raggedright
A joint CPTP map to $BR$, density operators for $R$,
tensor-product repetitions, and a quantum-measurement
representation of the recovery readout.}
\\

\parbox[t]{0.15\linewidth}{\raggedright
$2/3$ bound}
&
\parbox[t]{0.34\linewidth}{\raggedright
One allowed readout reproduces the probabilities of the
specified rank-one informationally complete input POVM.}
&
\parbox[t]{\dimexpr0.51\linewidth-4\tabcolsep\relax}{\raggedright
The complete experiment, including that readout, defines
a quantum instrument producing Bob’s qubit and the
readout’s classical outcome.}
\\

\parbox[t]{0.15\linewidth}{\raggedright
$1/2$ bound and approximate version}
&
\parbox[t]{0.34\linewidth}{\raggedright
An allowed recovery returns a qubit with
$F_{\mathrm{rec}}\geq1-\eta$; exact recovery has
$F_{\mathrm{rec}}=1$.}
&
\parbox[t]{\dimexpr0.51\linewidth-4\tabcolsep\relax}{\raggedright
The complete experiment, including recovery, defines
a joint CPTP channel producing Bob’s qubit and the
recovered qubit.}
\\

\hline\hline
\end{tabular}
\end{table}

% \begin{table}[htbp]
% \centering
% \caption{Recovery guarantees and operational descriptions assumed for the three bounds.}
% \label{tab:rob_assumptions}
% \small
% \setlength{\tabcolsep}{4pt}
% \renewcommand{\arraystretch}{1.15}

% \setlength{\robthirdcolwidth}{0.51\linewidth}
% \addtolength{\robthirdcolwidth}{-4\tabcolsep}

% \begin{tabular}{@{}p{0.15\linewidth}p{0.34\linewidth}p{\robthirdcolwidth}@{}}
% \hline\hline
% Bound & Recovery guarantee & Operational description assumed for that argument \\
% \hline
% Eventual tomography &
% An allowed estimator achieves $G_N>1/2$ at some finite $N$; eventual tomography guarantees that such a benchmark exists. &
% A joint CPTP map to $BR$, density operators for $R$, tensor-product repetitions, and a quantum-measurement representation of the recovery readout. \\
% $2/3$ bound &
% One allowed readout reproduces the probabilities of the specified rank-one informationally complete input POVM. &
% The complete experiment, including that readout, defines a quantum instrument producing Bob’s qubit and the readout’s classical outcome. \\
% $1/2$ bound and approximate version &
% An allowed recovery returns a qubit with $F_{\mathrm{rec}}\geq1-\eta$; exact recovery has $F_{\mathrm{rec}}=1$. &
% The complete experiment, including recovery, defines a joint CPTP channel producing Bob’s qubit and the recovered qubit. \\
% \hline\hline
% \end{tabular}
% \end{table}

\subsection{Copying accuracy and the scope of the witnesses}
\label{app:robustness_scope}

The copying assumption permits arbitrary operational accuracy in any fixed finite experiment.
For each bound, the implemented experiment must satisfy the operational assumptions of the corresponding argument.
For the tomographic bound, fix $N$ with $G_N>1/2$ before choosing that accuracy.
Let tildes denote quantities associated with the implemented copy-and-retain experiment, and suppose
\begin{equation}
    |\widetilde F_{\mathrm{tel}}-F_{\mathrm{tel}}|\leq\chi,
    \qquad
    |\widetilde G_N-G_N|\leq\chi.
    \label{eq:rob_copy_accuracy}
\end{equation}
Here $\widetilde G_N$ uses $\widetilde{\mathcal E}(\rho_\psi)^{\otimes N}$ with the same recovery measurement and guesses.
Thus $\chi$ bounds the change in the complete independent-record recovery score, not a per-copy error.
For $0\leq\chi<G_N-1/2$, the accuracy assumption ensures
$\widetilde G_N\geq G_N-\chi>1/2$, so the implemented
experiment retains an estimation advantage.
Applying Eq.~\eqref{eq:rob_finiteN_bound} to the implemented experiment,
and using the fidelity error estimate and the fact that
$f$ decreases, gives
\begin{equation}
\begin{aligned}
    F_{\mathrm{tel}}
    &\leq \widetilde F_{\mathrm{tel}}+\chi\\
    &\leq f\!\left(\frac{2\widetilde G_N-1}{N}\right)+\chi\\
    &\leq f\!\left(\frac{2G_N-2\chi-1}{N}\right)+\chi.
\end{aligned}
\label{eq:rob_copy_error}
\end{equation}
To obtain a fidelity ceiling strictly below one, $\chi$
must be small enough that the final right-hand side is
below one.
Such a choice is possible at fixed~$N$ because the ideal
bound is strictly below one and the expression is
continuous in~$\chi$.

For the measurement-reproduction bound, assigning the
guess $\ket{\chi_y}$ to outcome~$y$ gives ideal single-run
estimation fidelity $G=2/3$.
If copying changes both this score and the teleportation fidelity by at most~$\chi$, the
implemented readout satisfies
$\widetilde G\geq2/3-\chi$.
For $0\leq\chi\leq1/6$, substituting this lower bound
into the decreasing right-hand side of
Eq.~\eqref{eq:rob_banaszek} bounds
$\widetilde F_{\mathrm{tel}}$; adding~$\chi$ then bounds
$F_{\mathrm{tel}}$.

Similarly, if copying changes both the quantum-recovery
fidelity and the teleportation fidelity by at most~$\chi$, the guarantee
$F_{\mathrm{rec}}\geq1-\eta$ becomes
$\widetilde F_{\mathrm{rec}}\geq1-\eta-\chi$.
Equation~\eqref{eq:rob_approx_quantum_recovery} therefore
applies with $\eta$ replaced by $\eta+\chi$ and an
additive~$\chi$ on its right-hand side, provided
$\eta+\chi\leq1/2$.
Outside the stated parameter ranges, we retain the
trivial bound $F_{\mathrm{tel}}\leq1$.
Arbitrarily accurate copying recovers the ideal fidelity
ceilings by continuity at fixed~$N$; no exchange with an
infinite-sample limit is required.

These are necessary bounds, not claims that a fundamentally classical LMI mediator can attain the optimal quantum trade-offs.
A fidelity exceeding an applicable bound excludes a copyable mediator~\(C\), conditional on locality, the stated operational structure, and the relevant recovery guarantee.
The broader challenge is to extend these quantitative conclusions
beyond the operational descriptions assumed here: which weaker,
physically motivated assumptions can turn the main theorem's
general obstruction into an experimentally useful fidelity bound?

\end{document}